\documentclass[preprint,11pt,showpacs,nofootinbib]{revtex4}
\usepackage{graphicx}

\begin{document}

\title{Neutrino mean free paths in spin-polarized neutron Fermi liquids.}

\author{M. \'{A}ngeles P\'{e}rez-Garc\'{i}a}

\affiliation{Department of Fundamental Physics and IUFFyM,\\ University of Salamanca, Plaza de la Merced s/n 37008 Salamanca}

\date{\today}

\begin{abstract}
Neutrino mean free paths in magnetized neutron matter are calculated using the Hartree-Fock approximation with effective Skyrme and Gogny forces in the framework of the Landau Fermi Liquid Theory. It is shown that describing nuclear interaction with Skyrme forces and for magnetic field strengths $log_{10}\, B(G) \gtrsim 17$, the neutrino mean free paths stay almost unchanged at intermediate densities but they largely  increase at high densities when they are compared to the field-free case results. However the description with Gogny forces differs from the previous and mean free paths stay almonst unchanged or decrease at densities $[1-2]\rho_0$. This different behaviour can be explained due to the combination of common mild variation of the Landau parameters with both types of forces and the values of the nucleon effective mass and induced magnetization of matter under presence of a strong magnetic field as described with the two parametrizations of the nuclear interaction.
\vspace{1pc}
\pacs{21.30.Fe,21.65.Cd,26.60.-c.}
\end{abstract}

\maketitle

\section{Introduction}
\label{intro}

The study of the behaviour of hadronic matter in the density-temperature, $(\rho, T)$, diagram allows to have a deeper understanding of matter under extreme conditions. In this context, the high density, low temperature limit can be addressed for a fermion system using the Landau Fermi Liquid Theory (FLT) \cite{bookFL}. From a theoretical point of view the properties of this type of normal quantum systems can be studied calculating the interaction matrix element of quasiparticle (qp) excitations close to the Fermi surface. The inclusion of an additional component in the problem, a magnetic field, $B$, allows further testing the properties of magnetized Fermi Liquids. The role of magnetic fields in bulk properties and equation of state has been partially analyzed in the past for nuclear matter \cite{latt} \cite{chak} and quark matter \cite{quark1} \cite{quark2}.
Due to the tiny value of the neutron magnetic moment  $\mu_n=-1.9130427(5)\mu_N$ ($\mu_N=3.152 451 2326(45)$ $\times 10^{-18}$ MeV $G^{-1}$)~\cite{pdb} and in order to provide a sizable magnetization,  huge magnetic fields are needed. 

The only scenarios where we have indication of such intense fields are, first, from estimates of the background magnetic fields created in heavy-ion collisions like those at RHIC \cite{rhic} and, second, in a subgroup of pulsars called magnetars. For these astrophysical objects surface magnetic field strengths are of the order $B \approx 10^{15}$ G \cite{thom, lazzati}. Recent numerical simulations \cite{sim} of formation of proto-neutron stars show that the field configuration plays a significant role in the dynamics of the core if the initial magnetic field is large enough. In particular, in the rapid cooling of the newly formed neutron-rich object neutrino transport is  an important ingredient \cite{cooling}. However, some  of these simulations lack from accurate and consistent neutrino transport, missing the impact of magnetic fields in the microphysics input that affects the dynamics of the collapsing dense objects.

In most of the existing calculations of nuclear matter (either symmetric, pure neutron or beta-equilibrated) the effect due to the presence of strong magnetic fields and the consistently induced spin polarization are discarded in a first approximation. Either relativistic \cite{prakash, chakra} or effective approaches \cite{vida1} have been used to obtain some insight into the equation of state (EOS) or some structure properties \cite{latt} in presence of magnetic fields. These include a possible transition to a ferromagnetic state, although simulations using realistic potentials seem to prevent it \cite{ferro}. In general, a non-vanishing magnetization in a low temperature nuclear plasma \cite{angprc} produces a resolution of some degenerated observables as obtained in the context of the FLT \cite{ang2, ang3}. 

\section{Formalism}
\label{sec2}

In this work we are interested in the response of a spin-polarized pure neutron plasma to a weak neutrino probe. It can be seen \cite{prakash} that for the density range $\rho \le 4\rho_0$, where the quark deconfinement is not expected to take place, and for magnetic field strengths of maximum strength $B \approx 10^{18}$ G, allowed in principle by the scalar virial theorem, the neutral system is mostly neutrons. The maximum magnetic field strength we will consider is $B^* \approx 2 \times 10^4$ (as measured in units of the electron critical  field $B^*=B/B^c_e$  with $B^c_e=4.4\,\times \,10^{13}$ G) and the neutron fraction is $Y_n > 0.98$ \cite{prakash}. So the neutral plasma is mostly neutrons but leptons and additional baryons are also present in a tiny fraction that should be considered for full application in an astrophysical scenario where $\beta$-equilibrium holds. 

We are interested in exploring the effect of a strong magnetic field and the spin polarization of a pure neutron plasma through the structure functions, which provide information on density and spin density in-medium correlations. The homogeneous system under study is under the presence of an  internal magnetic field, ${\bf B}=B {\bf k}$ populated by species with paricle density $\rho_{\sigma}$, where $\sigma=\pm 1$ is the  spin $z$-projection. $\Delta=\frac{\rho_+ - \rho_-}{\rho}$ is the spin excess and $\rho=\rho_+ + \rho_-$ is the total particle density. For given thermodynamical conditions $\Delta$ is obtained by minimizing the Helmholtz free energy per particle, $f(\rho, T, B,\Delta )=\epsilon-\mu_n \Delta \rho B$, where $\epsilon$ is the energy per particle. Note that parallel (antiparallel) aligned magnetic moments (spins) are energetically favoured. 
We have considered an effective approach to describe the nuclear interaction using zero-range Skyrme forces~\cite{vautherin} with two of the most widely used parametrizations given by the Lyon group SLy4 and SLy7~\cite{chabanat1,chabanat2} and finite range Gogny  with D1P \cite{d1p} and D1S \cite{d1s} forces. All of them provide good values for binding of nuclei and also for neutron matter EOS.

In the context of the FLT the properties of non-magnetized systems at low temperature have been evaluated \cite{plbbackman} by calculating the qp matrix element around the Fermi surface where the only dependence is on fermionic densities and the qp scattering angle, $\theta$, involved. In the usual formalism, for the non-magnetized case the qp matrix element is written as a multipolar expansion in Legendre polinomials,

\begin{equation}
V_{ph}=\sum_{l=0}^{\infty} \big [ f_l + g_l {\bf \sigma_1 .\sigma_2} \big ] P_l ( cos\theta) ,
\label{elemento}
\end{equation}
$f_l$ and $g_l$ are the so-called Landau parameters of multipolarity $l$. In the more general case where any two possible spin orientations $(\sigma ,\sigma')$ are taken into account, the polarized qp matrix elements \cite{notes} \cite{ang2} are a crucial ingredient to compute the response functions to a a weakly interacting neutrino probe that excites a collective mode $(\omega, q)$ under the presence of a magnetic field $B$. The Lindhard function in the system, $\chi^{(\sigma ,\sigma')}(\omega, q)$, satisfies the Bethe-Salpeter equation and can be written in the dipolar ($l \le 1$) case in the random phase approximation (RPA) as a coupled system,
\begin{eqnarray}
\chi^{(\sigma ,\sigma')} &=& \chi_0^{(\sigma)} \delta(\sigma, \sigma') +  
\chi_0^{(\sigma)} \sum_{\sigma''=+,-} f_0^{(\sigma \sigma'')}
\chi^{(\sigma'' \sigma')}\nonumber \\
&& + \gamma_1^{(\sigma)} \sum_{\sigma''=+,-} f_1^{(\sigma, \sigma'')}
\Gamma^{(\sigma'', \sigma')},
\label{chil1}
\end{eqnarray}
\begin{eqnarray}
\Gamma^{(\sigma ,\sigma')} &=& \gamma_1^{(\sigma)} \delta(\sigma, \sigma') 
+ \gamma_1^{(\sigma)} \sum_{\sigma''=+,-} f_0^{(\sigma \sigma'')}
\chi^{(\sigma'' \sigma')}\nonumber \\
&&  + \gamma_2^{(\sigma)} \sum_{\sigma''=+,-} f_1^{(\sigma, \sigma'')}
\Gamma^{(\sigma'' \sigma')},
\label{chig2}
\end{eqnarray}
with the auxiliar definitions, $\Gamma^{(\sigma, \sigma')}=\int \frac{d^3 k}{(2 \pi)^3} cos (\theta)\, G^{(\sigma,\sigma')}$ and $\gamma_n^{(\sigma)}=\int \frac{d^3 k}{(2 \pi)^3} cos^n (\theta)\, G_{0}^{(\sigma)}$. Notice that the qp propagators $G_{0}^{(\sigma)}$ have been given in \cite{annals} and the expressions for the coefficients $\gamma_i^{(\sigma)}$ can be written \cite{notes} in the Landau limit as $\gamma_1^{(\sigma)}= \nu^{(\sigma)} \chi^{(\sigma)}_0$ and $\gamma_2^{(\sigma)}=\nu^{2 (\sigma)} \chi^{(\sigma)}_0-\frac{k_{F,\sigma} m^{*}_{\sigma}}{6 \pi^2}$ where $\nu^{(\sigma)}=\frac{m^{*}_{\sigma} \omega}{k_{F,\sigma} q }$.  
The qp effective mass in a magnetized system depends on the polarized dipolar coefficients \cite{bookFL},
\begin{equation} 
 m^*_{\sigma}/m=1+\frac{1}{3} N_{0 \sigma} \big [ f_1^{(\sigma,\sigma)}+(\frac{k^2_{F,-\sigma}} {k^2_{F,\sigma}})f_1^{(\sigma,-\sigma)} \big ] 
\label{mef}
\end{equation}
where  $N_{0 \sigma}=\frac{m^{*}_{\sigma} k_{F,\sigma}}{2 \pi^2}$ is the quasiparticle  level density  at each polarized Fermi surface with momentum $k_{F,\sigma}$. 

The generalized parameters $f_l^{(\sigma,\sigma')}$ are obtained by derivating the Helmhotz free energy with respect to the polarized density component, $f_{{\bf k},\sigma,{\bf k}',\sigma'}=\frac{\partial^2 F}{\partial n_{{\bf k},\sigma}\partial n_{{\bf k}',\sigma'}}$ \cite{ang2}, setting momenta on the polarized Fermi surfaces and expanding the resulting expression as a series in Legendre polinomials of  multipolarity $l$. These generalized parameters  fullfill the following relations recovering the usual ones in FLT in the limit $\Delta \rightarrow 0$ \cite{ang2},
\begin{equation}
f_l=\frac{f_l^{(\sigma,\sigma)}+f_l^{(\sigma,-\sigma)}}{2},
\label{f0sum}
\end{equation}
\begin{equation}
g_l=\frac{f_l^{(\sigma,\sigma)}-f_l^{(\sigma,-\sigma)}}{2}.
\label{g0sum}
\end{equation}
With the generalized paramters and using  the expressions in Eq. (\ref{chil1}) the corresponding Lindhard function for the isovector ($S=0$) response of the plasma can be written as,
\begin{equation}
\chi^{(S=0)} =  \chi^{(++)} + \chi^{(--)} + \chi^{(+-)} + \chi^{(-+)} ,
\end{equation}
and for the vector-axial ($S=1$) response as,
\begin{equation}
\chi^{(S=1)} =  \chi^{(++)} + \chi^{(--)} -\chi^{(+-)} - \chi^{(-+)} .
\end{equation}
Then the previous expression of the Lindhard function in RPA approximation \cite{rpa} include in-medium correlations at zero temperature. From them, one can obtain the structure functions given by,
\begin{equation}
S^{S=0,1}(\omega,q)=\frac{-1}{\pi} Im \,\chi^{S=0,1}(\omega,q).
\end{equation}
The structure function allows to calculate the non-relativistic differential cross section of neutrinos scattering off matter via neutral currents from \cite{peth}
\begin{equation} 
\frac{1}{V} \frac{d \sigma} {d \Omega d \omega} = \frac{G^2_F}{8 \pi^3} E'^2 
[ C_V^2 (1+ cos \theta) S^{0}(\omega,q)+ C_A^2 (3-cos \theta) S^{1}(\omega,q)]
\label{cs}
\end{equation}
where $E(E')$ is the incoming (outgoing) neutrino energy and $\vec{k}$ $(\vec{k'})$ is the neutrino incoming (outgoing) three-momentum. The transferred energy is $\omega=E-E'$ and the transferred three-momentum is $\vec{q}=\vec{k} -\vec{k'}$. The neutral current vector and axial vector charges are $C_V=1/2$ and $C_A=-g_a/2$ where $g_a=1.260$ \cite{pdb}. $G_F/(\hbar c)^3=1.166\,39(1) \times 10^{-5} GeV^{-2}$ is the Fermi coupling constant. Once the response has been evaluated it is straightforward to evaluate the neutrino mean free paths in the medium , $\lambda^{-1}=\int \frac{1}{V} \frac{d \sigma} {d \Omega d\omega} d \Omega d \omega$.

\section{Results}
\label{results}

\begin{figure}
\begin{minipage}[b]{0.5\linewidth} 
\centering
\includegraphics[angle=-90,scale=.6]{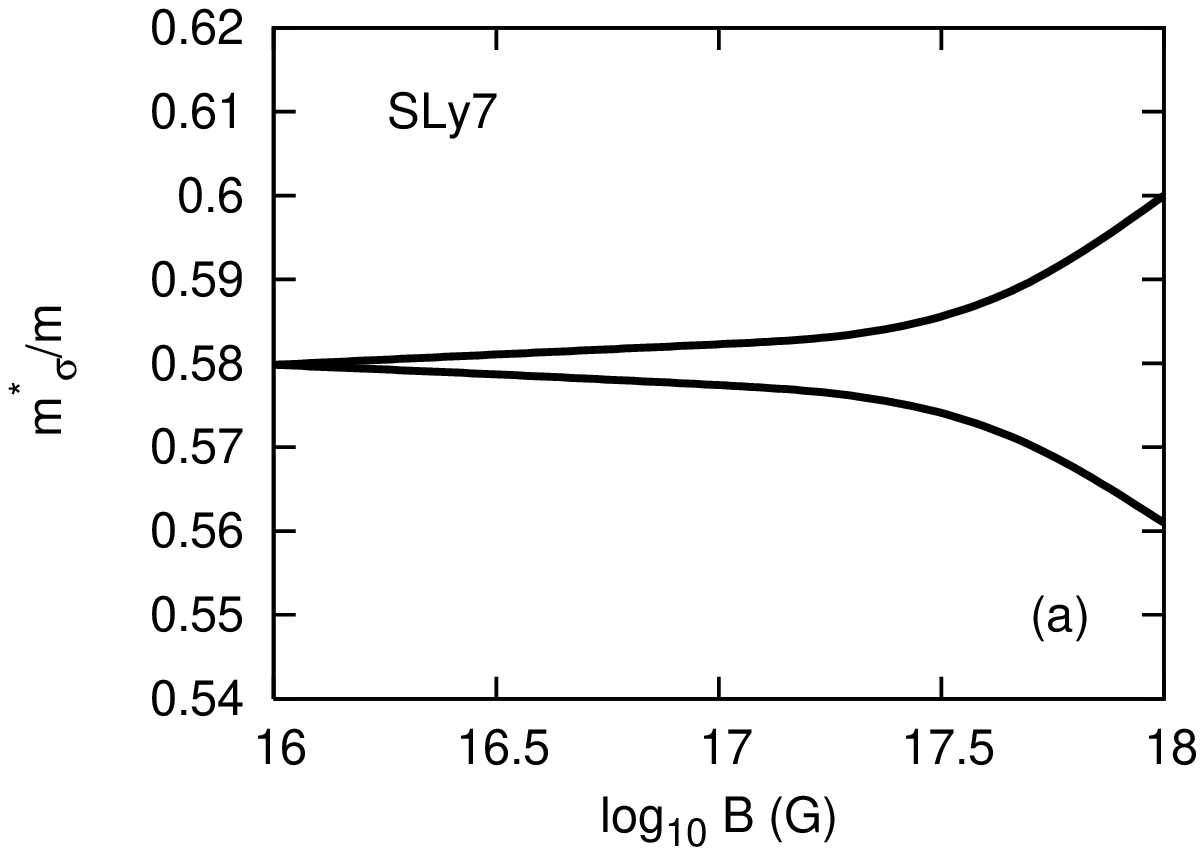}
\end{minipage}
\hspace{0.5cm} 
\begin{minipage}[b]{0.5\linewidth}
\centering
\includegraphics[angle=-90,scale=.6]{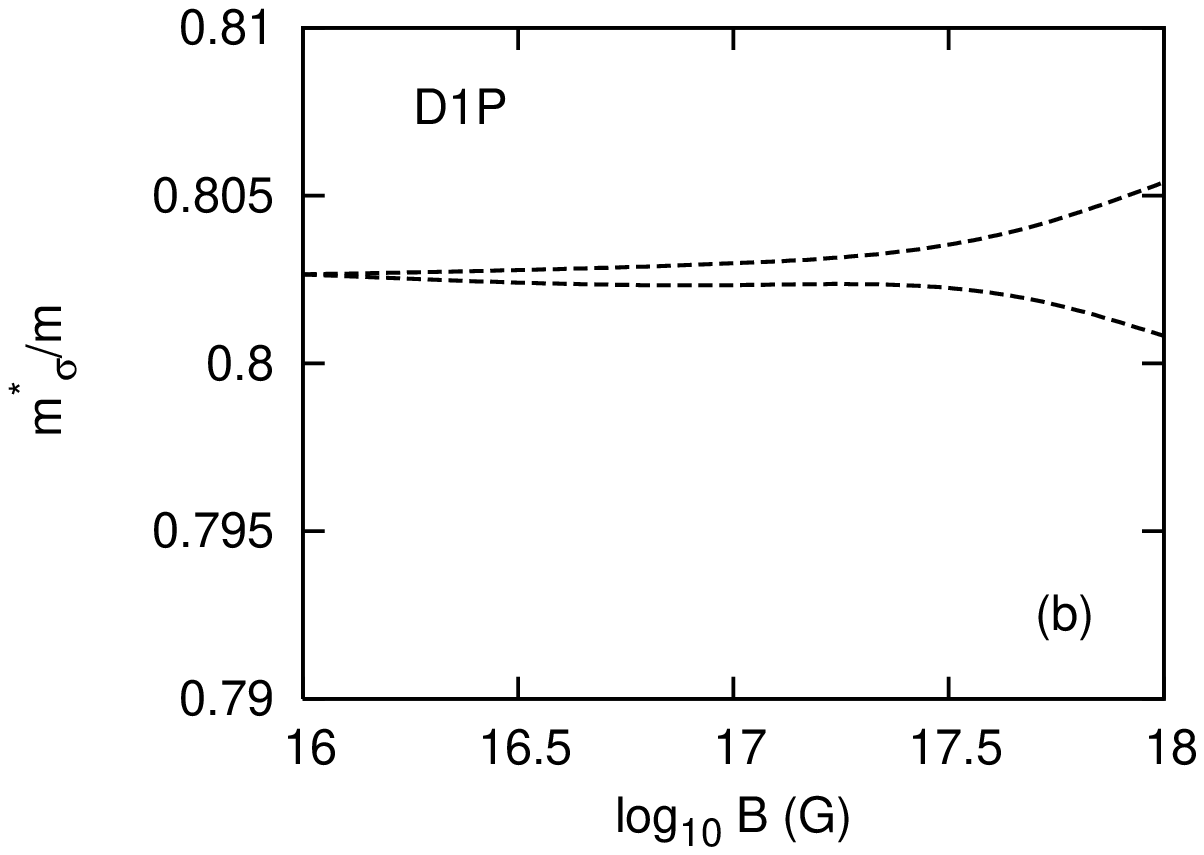}
\caption{Effective neutron mass at saturation density $\rho_0$ as a function of the logarithm of the magnetic field strength for the Skyrme SLy7 (a) and Gogny D1P (b) parametrizations. For each model upper (lower) curves correspond to spin up (down) polarized particles.} 
\label{Fig1}
\end{minipage}
\end{figure}

\begin{figure}[hbtp]
\begin{center}
\includegraphics [angle=-90,scale=.75] {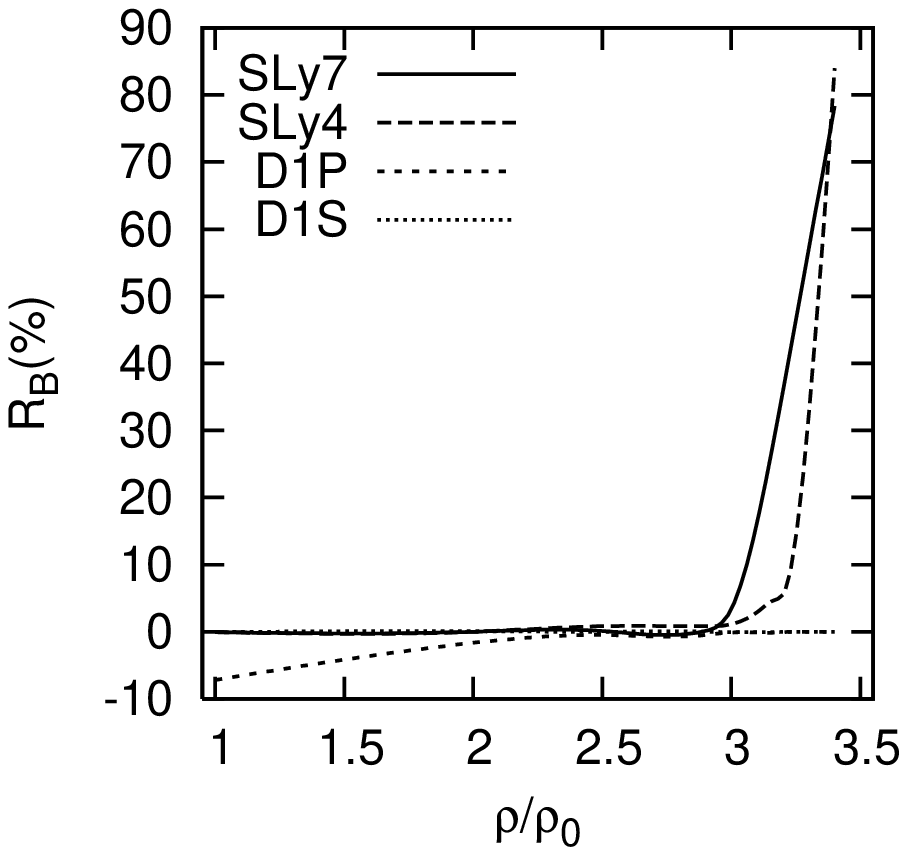}
\caption{Relative change ratio of neutrino mean free paths for $B=5 \times 10^{17}$ G with respect to the field-free case as a function of density calculated with Skyrme (SLy4 and SLy7) and Gogny (D1P and D1S) forces for a neutrino energy $E_{\nu}=15$ MeV.} 
\label{Fig2}
\end{center}
\end{figure}

In this section we include the effect of in-medium correlations in the neutron magnetized system as obtained in the Hartree-Fock approximation in the presence of a strong magnetic field. In Fig.~\ref{Fig1} the ratio of effective neutron mass as compared to the free value at saturation density, $\rho_0$, is shown as a function of the logarithm of the magnetic field strength for the Skyrme SLy7 (a) and Gogny D1P (b) parametrizations. For each model upper (lower) curves correspond to spin up (down) polarized particles. With Skyrme description the intense field affects more both effective nucleon mass absolute and relative (up-down polarized components) values than with Gogny. The impact of density effects on the mean free path can be seen in Fig.~\ref{Fig2}. We plot the ratio of change of neutrino mean free paths in the RPA dipolar approximation for a fixed value of magnetic field strength, $B=5 \times 10^{17}$ G, with respect to the field-free case, $R_B=\frac{\lambda_{B=5 \times 10^{17}G} - \lambda_{B=0}}{\lambda_{B=0}}$, as a function of the density (in units of nuclear  saturation density, $\rho_0$). We consider Skyrme SLy7 (solid line), SLy4 (long dashed line) Gogny D1P (short dashed line) and D1S (dotted line) parametrizations and set as a typical value of neutrino incoming energy $E_{\nu}=15$ MeV. While Gogny forces show almost unchanged or very mild reduction of mean free paths at densities in the range $[1-2]\rho_0$, the Skyrme forces show a high density dramatic increase with respect to the field-free case. Note that all standard Skyrme forces predict the onset of a ferromagnetic transition in the range $[1-4]\rho_0$, and in our selection of interactions for the study cases it is near $3.3\rho_0$. However this feature is not present in the Gogny forces that prevent ferromagnetic transitions. At densities $[1-2]\rho_0$, effects due to the energetic contribution of the magnetic perturbation, introduced by the neutron magnetic moment, are small at the selected field ($B=5 \times 10^{17}$ G) with respect to changes in other single particle properties like effective masses. For even lower densities (i.e. $0.5\rho_0$) it can be seen (see Fig. 7 in \cite{angprc}) that the $\mu_n \Delta \rho B$ term produces a relevant contribution to the magnetization. For application in astrophysical scenarios and at low densities one should consider that the effect of a non-zero proton fraction determines the appearance of pasta phases \cite{pasta} where electromagnetic and nuclear interactions are frustrated and clustering of matter arises. As density grows, at fixed values of $B$, the spin polarization decreases forming a {\it plateau} at intermediate densities before the possible appearance of a phase transition in the system. 

\begin{figure}[hbtp]
\begin{center}
\includegraphics [angle=-90,scale=.75] {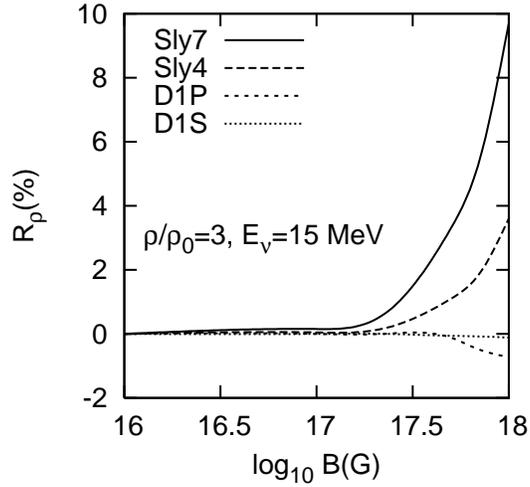}
\caption{Ratio of change of neutrino mean free paths as a function of the logarithm of magnetic field strength with Skyrme (SLy4 and SLy7) and Gogny (D1P and D1S) parametrizations at $\rho=3\rho_0$ and a neutrino energy $E_{\nu}=15$ MeV.} 
\label{Fig3}
\end{center}
\end{figure}

In  Fig.~\ref{Fig3} we plot the ratio of change of neutrino mean free paths in the RPA dipolar approximation computed for a generic  value of magnetic field strength with respect to the  field-free case as a function of the logarithm (base 10) of the magnetic field strength, $R_{\rho}=\frac{\lambda_B-\lambda_{B=0}}{\lambda_{B=0}}$. We set a value of density $\rho=3\rho_0$ and use SLy7 (solid line), SLy4 (long dashed line), D1P (short dashed line) and D1S (dotted line) parametrizations. For fields below $B \approx 10^{17}$ G there is almost no change in the ratio but for larger strengths there is a decrease (increase) as computed with Gogny (Skyrme) forces. For this high density case the change can be $\approx 10\%$ as computed with the SLy7 parametrization while the Gogny D1P predicts a relative change $\lesssim 1\%$. Note that the main contribution to the mean free paths comes from the fact that, as shown in Fig.~\ref{Fig1}, the Skyrme parametrization predicts a larger change in the absolute and relative values of the two effective masses of the spin polarized components. The Landau parameters and the energetic contribution of the magnetic perturbation \cite{ang2} show a minor contribution to the structure functions, that in turn determine the mean free paths. It is worth mentioning that the Lindhard function,$\chi^{(S)}$, has a rich structure in ($\omega, q$) that has been studied in \cite{ang3}, however, the smallnes of the magnetic perturbation is washed out in the response of the system by the influence of the magnetization \cite{angprc} and density effects in the neutron effective mass.  As we can see from Fig.~\ref{Fig3}, this result shows not only quantitative but also qualitative differences in the neutrino transparency of magnetized neutron matter. 
\section{Conclusions}
\label{conc}

We have investigated for the first time in the context of the Landau Theory of normal Fermi Liquids, the effect of a strong magnetic field on the variation of the neutrino mean free path in a partially magnetized pure neutron system  within the framework of the non-relativistic Hartree-Fock approximation comparing Skyrme and Gogny forces. We find that for fields up to the maximum strength studied in this work, $B=10^{18}$ G, Skyrme forces show at high density an enhancement of neutrino transparency of the system, while Gogny forces predict a small decrease. These results can be explanined due to the fact that for the density, and $B$ field range considered in this work the variation of Landau parameters is a minor contribution compared to the effective mass and magnetization.

\vspace{2ex}

\noindent{\bf Acknowledgments}\\
We acknowledge discussions with J. Navarro and A. Polls. This work has been partially funded by the Spanish Ministry of Science and Education under projects FIS2006-05319, FIS2009-07238 and Junta de Castilla y Leon Excellence program GR234. We also acknowledge support by CompStar, a research networking programme of the European Science Foundation.

\end{document}